# Nanoscale simulation of shale transport properties using the lattice Boltzmann method: permeability and diffusivity

Li Chen [a,b], Lei Zhang [b,c], Qinjun Kang [b], Jun Yao [c], Wenquan Tao [a]

a: Key Laboratory of Thermo-Fluid Science and Engineering of MOE, School of Energy and Power Engineering, Xi'an Jiaotong University, Xi'an, Shaanxi 710049, China

b: Earth and Environmental Sciences Division, Los Alamos National Laboratory, Los Alamos, New Mexico, USA

c: School of Petroleum Engineering, China University of Petroleum, Qingdao, Shandong 266580, China

Corresponding autor: Qinjun Kang: qkang@lanl.gov

**Abstract**

Porous structures of shales are reconstructed based on scanning electron microscopy (SEM) images of shale samples from Sichuan Basin, China. Characterization analyzes of the nanoscale reconstructed shales are performed, including porosity, pore size distribution, specific surface area and pore connectivity. The multiple-relaxation-time (MRT) lattice Boltzmann method (LBM) fluid flow model and single-relaxation-time (SRT) LBM diffusion model are adopted to simulate the fluid flow and Knudsen diffusion process within the reconstructed shales, respectively. Tortuosity, intrinsic permeability and effective Knudsen diffusivity are numerically predicted. The tortuosity is much higher than that commonly employed in Bruggeman equation. Correction of the intrinsic permeability by taking into consideration the contribution of Knudsen diffusion, which leads to the apparent permeability, is performed. The correction factor under different Knudsen number and pressure are estimated and compared with existing corrections reported in the literature. For the wide pressure range under investigation, the correction factor is always greater than 1, indicating the Knudsen diffusion always plays a role on the transport mechanisms of shale gas in shales studied in the present study. Most of the values of correction factor are located in the transition regime, with no Darcy flow regime observed.

**Keyword:** shale gas, Knudsen diffusivity, permeability, Klinkenberg correction, tortuosity, lattice Boltzmann method

Subject Areas: Fluid dynamics, Applied physics

Gas-bearing shale reservoirs have become major source of natural gas production in North American, and are expected to play increasingly important role in Europe and Asia in the near future. Shale is referred to extremely fine-grained sedimentary rocks with very small pores at the scale of nanometers [1]. Experimental observations using advanced techniques such as scanning electron microscopy (SEM) indicate that the components of shale include pores, organic matter (kerogen) and nonorganic minerals[2-4]. Natural gas trapped in shale formations is called shale gas. Shale gas includes free gas in pores and fractures as well as adsorbed gas on the surface of organic matters [1]. Shale gas is classified under unconventional gas because in shales nanoscale pores are dominant and porosity and permeability of shale are extremely low [1], leading to quite different transport mechanisms compared to that of natural gas in conventional reservoirs [1].

To extract shale gas from low porosity-permeability shale reservoirs more freely and economically, conductive pathways need to be artificially generated for shale gas transporting from the shale matrix to the wellbore. Currently, hydraulic fracturing is a widely used method to increase the permeability of a gas/shale formation by extending and/or widening existing fractures and creating new ones through the injection of a pressurized fluid into shale reservoirs [5]. With hydraulic fracturing, both shale matrix and fracture systems exist inside the shale gas reservoirs. Therefore, four levels of void space can be identified in shale reservoirs including mesoscale hydraulic fractures, meso/microscale natural fractures, micro/nanoscale interparticle pores and nanoscale kerogen pores [6,7]. Kerogen pores refer to nanoscale pores in the kerogen with the size ranging from 2nm-50nm [2,7].

A fundamental understanding of the multiscale multiple physiochemical processes during shale extraction is crucial for improving the gas production and lowering production costs. While the fracture network greatly determines the productivity of shale reservoirs, the transport of shale gas within the matrix also plays important roles. Due to the multiscale characteristics, different transport mechanisms play roles in different processes during the extraction of shale gas, including non-slip flow, slip flow, Knudsen diffusion, desorption-adsorption. Because of the ultra-low porosity and nanoscale characteristics of the shale matrix, Darcy law, which is widely adopted in the conventional formation, cannot realistically describe the variety of the relevant flow regimes other than the viscous flow regime [8]. The dominant transport mechanisms of shale gas in shale matrix is slip flow and Knudsen diffusion. Apparent permeability is suggested to use in the Darcy equation, which takes into account the effects of Knudsen diffusion and slip-flow, for the simulation of transport in shales. Several corrections have been proposed between the apparent and intrinsic permeability [8,9], and some of which are reported to be capable of describe all four flow regimes including continuum flow regime, slip flow regime, transition regime and free molecular regime.

To data, shale reservoirs and transport properties are still poorly understood, which is critical for understanding the fundamental transport mechanisms and evaluating the hydrocarbon recovery. There have been some researchers using experimental techniques to visualize the nanoscale structures of shale [1,2,7]. Due to the ultra-low porosity and permeability, it is difficult for experiment to fundamentally investigate the transport processes inside shales as well as accurately measure the transport properties. Alternatively, advanced pore-scale numerical simulation methods have been proved to be an efficient way [10-13], such as the lattice Boltzman method (LBM) [11,14]. In the present study, three-dimensional nanoscale porous structures of shales are reconstructed based on SEM images of shale samples. Structural characteristics of the reconstructed structures including porosity, pore size distribution, specific surface area and pore connectivity are analyzed in detail. The LBM is then applied to simulate fluid flow and diffusion processes in shales. Tortuosity, effective Knudsen diffusivity and intrinsic permeability of the reconstructed shales are numerically predicted. The ratio between apparent and intrinsic permeability is estimated and compared with existing corrections in the literature.

**Results**

Darcy equation is a constitutive equation that describes the flow of a fluid through a porous medium, which is expressed as

$$J_d = -\frac{\rho k_d}{\mu} \nabla p \qquad (1)$$

where $J_d$ (kg m$^{-2}$ s$^{-1}$) is the mass flux per unit area, $\mu$ (Pa.s) is viscosity, $\rho$ (kg m$^{-3}$) is density and $p$ (Pa) is pressure. Subscript "d" represents "Darcy". $k_d$ (m$^2$) is intrinsic permeability of the porous medium, which only depends on the porous structures and is not affected by the fluid type and flow condition. Darcy's law has been successfully applied to fluid flow in porous media in a wide range of scientific and engineering problems [15]. However, as the pore size decreases, or the Knudsen number $K_n$ (defined at the ratio between gas mean free path to the pore size) increases, the application of Darcy law is questionable, because the boundary slip phenomenon and the Knudsen diffusion begin to play important roles [16]. In nanoscale structures of shales, the transport behaviors of shale gas is a combined result of viscous flow and Knudsen diffusion[1,16,17]. Apparent permeability $k_a$, which can account for both transport mechanisms, is defined and used in Eq. (1)

$$J = -\frac{\rho k_a}{\mu}\nabla p, \quad J = J_d + J_k \tag{2}$$

where $J_d$ is given by Eq. (1). $J_k$ is the Knudsen diffusion flux term

$$J_k = -MD_{k,\text{eff}}\nabla C = -MD_{k,\text{eff}}\nabla(\frac{p}{zRT}) = -\frac{\rho}{p}D_{k,\text{eff}}\nabla p \tag{3}$$

where $M$ (kg mol$^{-1}$) and $C$ (mol m$^{-3}$) are molar mass and concentration, respectively. $R$ is the universal gas constant and $T$ (K) is temperature. $z$ is the gas compressibility factor, accounting for the effect of non-ideal gas. $D_{k,\text{eff}}$ (m$^2$ s$^{-1}$) is the effective Knudsen diffusivity. When the character length of a system is comparable to or smaller than the mean free path of the molecules, collisions between molecules and the solid wall are more frequent than that between molecules, and such diffusion is called Knudsen diffusion. The Knudsen diffusivity in a local pore with pore diameter $d_p$ (m) is calculated by [18]

$$D_k = \frac{d_p}{3}\sqrt{\frac{8RT}{\pi M}} \tag{4}$$

$d_p$ is an effective pore diameter calculated by the 13 direction averaging method proposed [19]. The effective Knudsen diffusivity $D_{k,\text{eff}}$ takes into account the pore size distribution and structural characteristics of a porous medium. Combining Eqs. (1), (2) and (3), the total flow flux can be expressed as

$$J = -k_d(1 + \frac{D_{k,\text{eff}}\mu}{pk_d})\frac{\rho}{\mu}\nabla p \tag{5}$$

Eq. (5) is the same to the single component gas transport model in a porous medium derived based on the dusty gas model (DGM) [14,20]. Shale gas is a mixture of several components, but the methane dominates with mole fraction of about 90 % [1]. Therefore, in this study shale gas is considered to be composed of pure methane, and thus binary gas diffusion is neglected [16]. Based on Eq. (5), the apparent permeability can be determined

$$k_a = k_d f_c, \quad f_c = (1 + \frac{D_{k,eff} \mu}{p k_d}) \qquad (6)$$

with $f_c$ is the correction factor [9,21]. Different expressions of $f_c$ have been proposed in the literature. Klinkenberg proposed the following first-order expressions [9]

$$f_c = 1 + 4cKn, \quad Kn = \frac{\lambda}{r} \qquad (7)$$

with $c \approx 1$. $Kn$ is the Knudsen number defined as the ratio between mean-free-path $\lambda$ and characteristic length $r$. Several second-order-corrections have been proposed [8,22]. Beskok and Karniadakis [22] developed the following correction

$$f_c = (1 + \alpha(Kn)Kn)\left[1 + \frac{4Kn}{1 - bKn}\right] \qquad (8)$$

with $b$ is the slip coefficient and equal to -1 for slip flow. $\alpha(Kn)$ is the rarefaction coefficient, the expression of which is very complex in [22]. Civan [8] developed a much simplified $\alpha(Kn)$

$$\alpha(Kn) = \frac{1.358}{1 + 0.170 Kn^{-0.4348}} \qquad (9)$$

Eqs. (8) and Eq. (9) are called Beskok and Karniadakis-Civan's correction in the present study.

Here, the correction factor predicted by Eq. (6) is compared with that by Klinkenberg's correction (Eq. (7)) and Beskok and Karniadakis-Civan's correction (Eqs. (8) and (9)). A simple case of fluid flow and Knudsen diffusion in a cylinder with diameter $d_p$ are considered. For such structure, the intrinsic permeability is

$$k_d = \frac{(d_p / 2)^2}{8} \qquad (10)$$

With the mean-free-path $\lambda$ given by [8]

$$\lambda = \frac{\mu}{p} \sqrt{\frac{\pi RT}{2M}} \qquad (11)$$

Substituting Eqs. (4) and (10) into Eq. (6), leads to

$$f_c = (1 + \frac{64}{3\pi} Kn) \tag{12}$$

Fig. 1 shows the correction factor in the cylinder predicted by different corrections. When $Kn$ is quite low, i.e., in the range of Darcy flow regime ($Kn<0.01$), the above equations predict quite similar values as the term with $Kn$ can be neglected. However, as $Kn$ increases, the discrepancy between Klinkenberg's correction and other corrections becomes large. The Beskok and Karniadakis-Civan's correction is more accurate than Klinkenberg's correction, especially at high $Kn$ regime [21]. Values predicted by Eq. (12) agree well with Beskok and Karniadakis-Civan's correction. This agreement leads to the theoretical basis of the present study.

In the literature, for determine the apparent permeability of microchannels or micro porous media, in which the fluid flow regime falls in slip flow or transition flow regime, Navier-stokes equation is solved with modified local viscosity and modified slip flow boundary conditions [23,24]. Delicate numerical schemes are required in such simulations. Through such simulations, the detailed distribution of fluid flow, pressure as well as the apparent permeability can be obtained. However, for reservoir simulations, the empirical relationships (permeability-porosity, diffusivity-porosity, etc.) rather than the detailed transport information are more required. Therefore, the analysis related to Fig. 1 indicates that one can predict the intrinsic permeability and the Knudsen diffusivity of a porous medium, and then use Eq. (6) to determine the apparent permeability. Such a scheme avoid the tedious numerical schemes required for accurately predict the detailed fluid flow filed in microchannels or micro porous media [23,24].

The first step towards performing the nanoscale simulations based on LBM is to obtain the digitalized structures of shales. The shale sample used to generate the SEM images comes from a well located in the Sichuan Basin, china [25]. SEM is adopted to scan different parts of the shale sample and 13 SEM images are generated (one of them is shown in Fig. 2(a)) [25]. Resolution of the SEM image is $\delta x=5nm$. Using the three-dimensional (3D) reconstruction technique based on Markov chain Monte Carlo (MCMC) (For more details of the reconstruction, one can refer to [25]), four SEM images are selected to reconstruct the 3D nanoscale porous structures of the shale. Fig. 2(b) shows one of the reconstructed structures. The size of each domain $L_x \times L_y \times L_z$ is 250×100×100 lattices with physical size of 1250nm×500nm×500nm.

Using the numerical method introduced in Methods Section, the Knudsen diffusion and fluid flow inside the shale matrix are simulated. Fig. 3 shows the 3D distribution of methane concentration in the reconstructed structures obtained from the LBM mass diffusion model. Fig. 4 shows the effects of the temperature on the effective diffusivity. Fig. 5 displays the 3D distribution of pressure and streamline in shales. Based on the concentration and velocity distribution, the Knudsen diffusivity and intrinsic permeability are predicted. The apparent permeability is then calcualted based on Eq. (6) and are compared with existing empirical corrections under different pressure (Fig. (6)) and different $Kn$ (Fig. (7)).

**Discussion**

**Structure characterization.** Characterization analysis of the porous structures provides necessary information for the predictions of the transport properties of the shale. Porosities of the four structures are 0.191, 0.226, 0.268 and 0.176, respectively, which are close to that in the literature [2,6,11]. Pore size distribution (PSD) is an important structure characteristic of porous media. Determining the pore size of each local pore cell is also required for the calculating of the local Knudsen diffusivity as shown in Eq. (4). The 13 direction averaging method proposed in Ref. [19] is adopted to calculate the pore size of each pore cell. Fig. 2(c) shows the PSD of the reconstructed structures shown in Fig. 2(c), which presents the unimodal distributions. For all the reconstructed structures, the pore size is in the range 5~80 nm, indicating the nanoscale characteristics of shales.

Specific surface area is defined as the ratio between the total surface area and the total volume of the solid phase. Although gas desorption is a relatively slow process, it can account up to 50% of the total gas production [26]. Desorption process is significantly affected by the specific surface area of shales. For the structures shown in Fig. 2(b), the total surface area and total volume of solid phase is 721438 $(\delta x)^2$ and 2043351 $(\delta x)^3$, respectively, leading to specific surface area as 0.353066$(\delta x)^{-1}$. With $\delta x$=5nm, the specific surface area as $7.06 \times 10^7$ m$^{-1}$ and 47.07 m$^2$ g$^{-1}$, with kerogen density as 1.5 g cm$^{-3}$ [27]. The specific surface area in the present study is higher than 14.0 m$^2$ g$^{-1}$ estimated in [11], due to high resolution in the present study compared to 12 nm in [11]. The specific surface area for other reconstructed structures is 46.72, 85.71 and 90.08 m$^2$ g$^{-1}$, respectively.

Shale gas transports through the connected pores in the organic matter. Connectivity of the void space is therefore very important. For the reconstructed structures of kerogen, "transport" and "dead" pore cells are distinguished. Here, a pore cell is "transport" means that this cell belongs to a continuous percolation path throughout the entire kerogen. If a portion of void phase does not penetrate the entire kerogen, shale gas entering the shale from one end cannot arrive at the other end through this portion, and this portion is called "dead". For identifying the "transport" and "dead" portions in the reconstructed kerogen, a connected phase labeling algorithm is developed [28]. This algorithm scans the entire domain, checks the connectivity of a cell with its neighboring 18 points, labels the cell depending on the local connectivity, and finally assigns any portion with a distinct label if it is disconnected from other portions in the domain. Here, neighboring 18 points are used, in coincidence with the D3Q19 lattice model in the LBM framework adopted in the present study. The connectivity of void space is defined as the ratio of the number of "transport" cells to the total pore cells. Fig. 2(d) shows the void phase cells after labeled. The blue portion represents the "transport" one while the red portions inside the domain denote the "dead" portions. For Sample 4, simulation results show that there are 358 portions of void space in the reconstructed CL with one of them as "transport" and the remaining as "dead". The "transport" portion occupies most of the void phase, with a volume fraction of 98.0% (447460/456649, 456649 is the total number of pore cells), as clearly seen in Fig. 2(d) where the "dead" portions present as small and discrete blobs. The connectivity for other reconstructed structures is 99.1%, 99.7% and 99.8%, respectively.

**Effective Knudsen diffusivity.** As shown in Fig. 3, the concentration distribution is very complicated and is greatly affected by the local pore space. The effective Knudsen diffusivity is calculated based on the concentration distribution shown. The ratio between $D_{k,eff}$ and $D_0$ is 0.0212, 0.0197, 0.0380 and 0.0130 for Samples 1-4, respectively. $D_0$ is chosen as the Knudsen diffusivity with $d_p$=25nm. With $T$=383K, $D_0$=5.44×10$^{-6}$ m$^2$/s. Therefore, $D_{k,eff}$ for Samples 1-4 is 1.25×10$^{-7}$ m$^2$/s, 1.17×10$^{-7}$ m$^2$/s, 2.25×10$^{-7}$ m$^2$/s and 7.71×10$^{-8}$ m$^2$/s, respectively. According to Eq. (4), the temperature will affect the Knudsen diffusivity. Fig. 4 shows the variation of effective Knudsen diffusivity with temperature. $D_{k,eff}$ increases as the temperature increases. For temperature with a change of about 100K, the variation of $D_{k,eff}$ is not quite significant.

Macroscopic models of transport processes in shale matrix highly depend on empirical relationships between macroscopic transport properties and statistical structural information of porous components (permeability VS porosity, diffusivity VS porosity. Bruggeman equation has been widely used in the macroscopic models to calculate the effective diffusivity

$$D_{eff} = D_0 \frac{\varepsilon}{\tau} \quad (13)$$

$\tau$ is the tortuosity of a porous medium. It is defined as the ratio between the actual flow path length and the thickness of a porous medium along the flow direction. Higher tortuosity indicates longer, more tortuous and sinuous paths, thus resulting in greater transport resistance. In Bruggeman equation, the tortuosity is commonly defined as an exponential function of porosity

$$\tau = \varepsilon^{-\alpha} \quad (14)$$

with α usually chosen as 0.5. Bruggeman equation has been widely used in the mascroscopic models to calculate the effective transport properties. However, it was determined empirically from sphere packed porous medium, which therefore cannot reflect the complex structures of shales. Based on the predicted diffusivity, Eq. (13) is adopted to calculate $\tau$, and Eq. (14) is then used to determine α. α determined based on the simulation results ranges 1.33~1.65, much higher than 0.5 in the Bruggeman equation. Pore-scale modeling, which is based on the realistic porous structures of shales, fundamentally reflects the actual transport processes inside shales. Therefore, the simulation results indicate that the pathway inside the shales is much tortuous. It can be concluded that the Bruggeman equation overestimates the effective diffusivity, and higher α needs to be adopted in the Bruggeman equation to more accurately calculate effective diffusivity in shales. In fact, for complex nanoscale porous medium, such as catalyst layer of proton exchange membrane fuel cell, the value of α is often found to be much higher than that used in the Bruggeman equation, where α in the range of 1.0~2.0 is predicted [19].

**Intrinsic permeability.** Fluid flow within the reconstructed shales is simulated using the LBM fluid flow model. The pressure difference between the inlet and outlet, which is expressed by the density difference in the simulation, is 0.0002, leading to the maximum magnitude of velocity about $4\times10^{-6}$ in lattice units, meeting the low March number limit of LB fluid flow model. As shown in Fig. 5, the streamlines are quite tortuous, consistent with the high tortuosity predicted. The continuous pathways from the inlet to the outlet are quite few, resulting in quite low permeability of shales. With $\delta x$=5nm, the permeability for Samples 1~4 is 240.5nD, 172.75nD, 367.5nD, 135.0nD, respectively, with 1nD=$10^{-21}$ m$^2$. Javadpour experimentally measures 152 samples from nine reservoirs and found that the 90% o f the measured permeability is less than 150 nD [1]. Our predicted results are a little high due to the relative high porosity. Chen et al. [11] also performed LB simulations of a reconstructed shale with porosity of 29.9%, and the permeability predicted were about 1500~3920 nD in different directions. According to KC equation [29], the higher the porosity, the larger the permeability. The reverse trend of $D_{k,eff}/D_0$ and $k_d$ for Sample 1 and 2 indicates the complexity of the porous structures of shales.

**The apparent permeability.** Values of the correction factor are predicted based on the intrinsic permeability and effective Knudsen diffusivity obtained from the nanoscale simulations, and are compared with that predicted by Klinkenberg's correction (Eq. (7)) and Beskok and Karniadakis-Civan's correction (Eqs. (8) and (9)). Pressure is chosen in the range of 10-4000 psi (1psi=6894.75729 Pa). The temperature is fixed at 383K. Viscosity of methane is calculated using a online software called Peace Software [30]. In Eqs. (7), (8) and (9), the following equation derived by Civan is used to calculate $Kn$ [8]

$$Kn = \frac{\mu}{4}\sqrt{\frac{\pi RT}{\tau M}}(\frac{k_d}{\varepsilon})^{-0.5} \tag{15}$$

based on the intrinsic permeability, porosity and tortuosity of the samples predicted by our LBM simulations.

Fig. 6 shows the comparison between the correction factor $f_c$ predicted by the LBM simulation and that predicted by Eqs. (7) and (8-9) under different pressure, and Fig. 7 shows the comparison under different $Kn$. For the wide pressure range under investigation, $f_c$ is always greater than 1, indicating the Knudsen diffusion always plays a role on the transport mechanisms of shale gas in shales studied in the present study. This is also confirmed in Fig. 7, where all data falls within the slip flow regime (0.01<$Kn$<0.1), transition flow regime (0.1<$Kn$<10) and Knudsen diffusion regime ($Kn$>10), with no Darcy flow ($Kn$<0.01) regime observed as $Kn$ is always greater than 0.03. As the pressure decreases or the $Kn$ increases, the Knudsen diffusion becomes increasingly dominant. For pressure lower than 100 psi, the correction factor is higher than 10, meaning the Knudsen diffusion dominates and the viscous flow can be ignored. Most of the values of correction factor are located in the transition regime. Besides, the simulation results are in better agreement with that predicted by Beskok and Karniadakis-Civan's correction compared with that of Kinkenberg's correction, in consistence with that in Fig. 1. The values from our simulations are a little higher than that of Beskok and Karniadakis-Civan's correction. Ziarani and Aguileran [21] also found Beskok and Karniadakis-Civan's correction underestimates the correction factor, compared with the experimental data from Mesaverde formation. Note that the Beskok and Karniadakis-Civan's correction as well as Eq. (4) are derived based on the diffusion process in a cylinder [18,22]. However, pores in the shales are complex and can hardly be described by a collection of cylinders. Using Eq. (4) will overestimate the Knudsen diffusivity in porous media. Therefore, in the future simulations, complex structures of the pores as well as the nature of the redirecting collision between walls and molecules should be taken into account by improving Eq. (4) [31,32].

## Methods

Owing to its excellent numerical stability and constitutive versatility, the LBM has developed into a powerful technique for simulating transport processes and is particularly successful in transport processes applications involving interfacial dynamics and complex geometries, such as multiphase flow and porous media flow [33-35]. In the present study, intrinsic permeability and effective Knudsen diffusivity of the reconstructed shales are numerically predicted using the Multiple-relaxation-time LB fluid flow model and LB mass transport model, respectively.

**Multiple-relaxation-time LB fluid flow model**. A viscosity-dependent permeability is usually obtained when adopting SRT LBM for simulating fluid flow [36]. In order to overcome such defect, the multi-relaxation-time (MRT) model has been proposed recently to simulate fluid flow through porous media [36,37]. The MRT-LBM model transforms the distribution functions in the velocity space of the SRT-LBM model to the moment space by adopting a transformation matrix. In the SRT-LBM model, the evolution equation for the distribution functions is as follows

$$f_i(\mathbf{x}+\mathbf{e}_i\Delta t, t+\Delta t) - f_i(\mathbf{x},t) = \mathbf{S}[f_i^{eq}(\mathbf{x},t) - f_i(\mathbf{x},t)] \quad i = 0 \sim N \tag{16}$$

where $f_i(\mathbf{x},t)$ is the $i$th density distribution function at the lattice site $\mathbf{x}$ and time $t$. $\mathbf{S}$ is the relaxation matrix. For the D3Q19 (three-dimensional nineteen-velocity) lattice model with $N=18$ used in this work as shown in Fig. 5(a), the discrete lattice velocity $\mathbf{e}_i$ is given by

$$\mathbf{c}_i = \begin{cases} 0 & i=0 \\ (\pm 1,0,0),(0,\pm 1,0),(0,\pm 1,0) & i=1\sim 6 \\ (\pm 1,\pm 1,0),(0,\pm 1,\pm 1),(\pm 1,0,\pm 1) & i=7-18 \end{cases} \quad (17)$$

$f^{eq}$ is the $i$th equilibrium distribution function and is a function of local density and velocity

$$f_i^{eq} = w_i \rho \left[ 1 + \frac{\mathbf{e}_i \cdot \mathbf{u}}{(c_s)^2} + \frac{(\mathbf{e}_i \cdot \mathbf{u})^2}{2(c_s)^4} - \frac{\mathbf{u} \cdot \mathbf{u}}{2(c_s)^2} \right] \quad (18)$$

with the weight coefficient $w_i$ as $w_i=1/3$, $i=0$; $w_i=1/18$, $i=1,2,\ldots, 6$; $w_i=1/36$, $i=7,8,\ldots,18$. $c_s = 1/\sqrt{3}$ is the speed of sound. By multiplying a transformation matrix $\mathbf{Q}$ (a $(N+1)\times(N+1)$ matrix) in Eq. (16), the evolution equation in the moment space can be expressed as

$$\mathbf{m}(\mathbf{x}+\mathbf{c}\Delta t, t+\Delta t) - \mathbf{m}(\mathbf{x},t) = \hat{\mathbf{S}}[\mathbf{m}^{eq}(\mathbf{x},t) - \mathbf{m}(\mathbf{x},t)] \quad (19)$$

where

$$\mathbf{m} = \mathbf{Q} \cdot \mathbf{f}, \quad \mathbf{m}^{eq} = \mathbf{Q} \cdot \mathbf{f}^{eq}, \quad \hat{\mathbf{S}} = \mathbf{Q} \cdot \mathbf{S} \cdot \mathbf{Q}^{-1} \quad (20)$$

with $\mathbf{m}$ and $\mathbf{m}^{eq}$ as the velocity moments and equilibrium velocity moments, respectively. $\mathbf{Q}^{-1}$ is the inverse matrix of $\mathbf{Q}$ [44]. The transformation matrix $\mathbf{Q}$ is constructed based on the principle that the relaxation matrix $\hat{\mathbf{S}}$ (a $(N+1)\times(N+1)$ matrix) in moment space can be reduced to the diagonal matrix [37], namely

$$\hat{\mathbf{S}} = \text{diag}(s_0, s_1, \ldots, s_{17}, s_{18}) \quad (21)$$

with [36]

$$s_0 = s_3 = s_5 = s_7 = 0, \quad s_1 = s_2 = s_{9-15} = \frac{1}{\tau}, \quad s_4 = s_6 = s_8 = s_{16-18} = 8\frac{2\tau - 1}{8\tau - 1} \quad (22)$$

$\tau$ is related to the fluid viscosity by

$$\tau = \frac{\upsilon}{c_s^2 \Delta t} + 0.5 \quad (23)$$

The equilibrium velocity moments $\mathbf{m}^{eq}$ are as follows [36]

$$\mathbf{m}_0^{eq} = \rho \tag{24a}$$

$$\mathbf{m}_1^{eq} = -11\rho + 19\frac{\mathbf{j} \cdot \mathbf{j}}{\rho_0}, \quad \mathbf{m}_2^{eq} = 3\rho - \frac{11}{2}\frac{\mathbf{j} \cdot \mathbf{j}}{\rho_0} \tag{24b}$$

$$\mathbf{m}_3^{eq} = j_x, \quad \mathbf{m}_4^{eq} = -\frac{2}{3}j_x \tag{24c}$$

$$\mathbf{m}_5^{eq} = j_y, \quad \mathbf{m}_6^{eq} = -\frac{2}{3}j_y \tag{24d}$$

$$\mathbf{m}_7^{eq} = j_z, \quad \mathbf{m}_8^{eq} = -\frac{2}{3}j_z \tag{24e}$$

$$\mathbf{m}_9^{eq} = \frac{3j_x^2 - \mathbf{j} \cdot \mathbf{j}}{\rho_0}, \quad \mathbf{m}_{10}^{eq} = -\frac{3j_x^2 - \mathbf{j} \cdot \mathbf{j}}{2\rho_0} \tag{24f}$$

$$\mathbf{m}_{11}^{eq} = \frac{j_y^2 - j_z^2}{\rho_0}, \quad \mathbf{m}_{12}^{eq} = -\frac{j_y^2 - j_z^2}{2\rho_0} \tag{24g}$$

$$\mathbf{m}_{13}^{eq} = \frac{j_x j_y}{\rho_0}, \quad \mathbf{m}_{14}^{eq} = \frac{j_y j_z}{\rho_0}, \quad \mathbf{m}_{15}^{eq} = \frac{j_x j_z}{\rho_0} \tag{24h}$$

$$\mathbf{m}_{16-18}^{eq} = 0 \tag{24i}$$

where density and momentum are determined by

$$\rho = \sum_i f_i, \quad \mathbf{j} = \sum_i f_i \mathbf{e}_i \tag{25}$$

$\rho_0$ in Eq. (24) is the mean density of the fluid, which is employed to reduce the compressibility effects of the model [36,37]. Eqs. (19) and (24) can be recovered to Navier-Stokes equations using Chapman–Enskog multiscale expansion under the low Mach number limitation.

**SRT LB mass transport model.** In this work, for the first time the LBM is adopted to simulate the Knudsen diffusion in shales. Since SRT LB mass transport model is sufficient to accurately predict the pure diffusion process, it is employed rather than MRT LB one. For pure methane

diffusion in shales, the evolution equation for the concentration distribution function is as follows

$$g_i(\mathbf{x}+\mathbf{e}_i\Delta t, t+\Delta t) - g_i(\mathbf{x},t) = -\frac{1}{\tau_g}(g_i(\mathbf{x},t) - g_i^{eq}(\mathbf{x},t)) \qquad (26)$$

where $g_i$ is the distribution function. It is worth mentioning that for simple geometries, D3Q7 lattice model (or D2Q5 model in 2D) is sufficient to accurately predict the diffusion process and properties, which can greatly reduce the computational resources, compared with D3Q19 (or D2Q9 in 2D), as proven by our previous work [38,39]. However, for complex porous structures, especially for those with low porosity, such as shales, using reduced lattice model will damage the connectivity of the void space, thus leading to underestimated effective diffusivity. Therefore, in coincidence with the 19 direction labeling algorithm introduced in Section 3.4, D3Q19 lattice model is adopted. The equilibrium distribution function $g^{eq}$ is defined

$$g_i^{eq} = C_{CH_4}/a_i, \quad a_i = \begin{cases} 1/3 & i=0 \\ 1/18 & i=1\sim 6 \\ 1/36 & i=7-18 \end{cases} \qquad (27)$$

The concentration and the diffusivity are obtained by

$$C_{CH_4} = \sum g_i, \quad D = \frac{1}{3}(\tau_g - 0.5)\frac{\delta x^2}{\delta t} \qquad (28)$$

Eqs. (26) and (27) can be proved to recover the pure diffusion equation using Chapman–Enskog multiscale expansion [39]. According to Eq. (4), the Knudsen diffusivity is a function of local pore diameter. Based on the PSD of the reconstructed shales, a reference pore size $d_{p,ref}$=25 nm is chosen, and the corresponding relaxation time $\tau_{g,ref}$ is set as 1.0. Hence, for any local pore with size $d_p$, the relaxtion time can be calculated as

$$\tau_g = \frac{d_p}{d_{p,ref}}(\tau_{g,ref} - 0.5) + 0.5 \qquad (29)$$

Through Eq. (29), the effects of pore size distributions on the Knudsen diffusion process inside the nanoscale structures of porous shales are taken into account.

**Boundary condition.** For MRT-LB fluid flow model, the collision step is implemented in the moment space while the stream step is carried out in the velocity space. Therefore, the boundary conditions developed in SRT model can still be applied to MRT-LBM. In this study, for fluid flow three kinds of boundary conditions are adopted: no-slip boundary condition at the fluid-solid interface inside the domain, pressure boundary condition at the inlet and outlet of the computational domain ($x$ direction), and periodic boundary conditions for the other four boundaries ($y$ and $z$ directions). In the LB framework, for no slip boundary condition, the half way bounce-back scheme is employed; for periodic boundary conditions, the unknown distributions at one boundary ($y=0$ for example) is set to that at the other boundary ($y=L_y$, for example); and for pressure boundary condition, the non-equilibrium extrapolation method proposed by Guo et al. [40] is adopted for its good accuracy. It should be noted that the pressure difference between inlet and outlet should be small enough, one reason to ensure low Mach number fluid flow for the LB model and the other reason to eliminate the inertial effect.

For methane diffusion, four types of boundary conditions are used: the no flux boundary condition at the fluid-solid interface inside the domain, concentration boundary condition at the inlet and outlet of the computational domain ($x$ direction), and periodic boundary conditions for the other four boundaries ($y$ and $z$ directions). In the LB framework, for the concentration boundary condition, the unknown distribution function is determined using the equilibrium distribution functions in Eq. (27); for the no flux boundary condition, bounce-back scheme is adopted.

**Macroscopic transport properties.** When the simulation of fluid flow reaches steady state using the MRT-LBM fluid flow model, the intrinsic permeability is calculated using Eq. (1), where the flow flux $J_d$ is expresses as

$$J_d = \rho \langle u \rangle \quad (30)$$

with $<u>$ as the volume-averaged fluid velocity along flow direction.

The effective Knudsen diffusivity $D_{K,eff}$ along the $x$ direction is calculated by

$$D_{K,\text{eff}} = \frac{(\int_0^{L_y}\int_0^{L_z}(D_K \frac{\partial C}{\partial x})|_{L_x} dydz)/L_y L_z}{(C_{\text{in}} - C_{\text{out}})/L_x} \tag{31}$$

where $C_{\text{in}}$ and $C_{\text{out}}$ is the inlet and outlet concentration, respectively. $|_{L_x}$ means local value at $x=L_x$.

**Validation.** For validation of the MRT-LBM fluid flow model, simulation is performed for flow through an 3D open cube in which equal-sized spheres of diameter $d$ are arranged in periodic BCC (body-centered cubic) arrays and $Re$ number is much less than unity, as shown in Fig. 8. 100×100×100 lattice is employed, with pressure difference applied on the left and right boundaries and periodic boundary conditions on the other four directions. The insert of Fig. 8 shows the simulated streamlines inside the periodic BCC arrays with a porosity of 0.915. A comparison between the simulated permeability and Kozeny-Carman (KC) equation [29] for the BCC arrays with different porosity is shown in Fig. 8. The KC equation, a widely used semi-empirical equation for predicting permeability, is given by $k_d = A\varepsilon^3/(1-\varepsilon)^2$, with $A$ is the KC constant and is set as $C = (d^2/180)$ for packed-spheres porous media. Analytic solutions [41] are also displayed in Figure. 8. As shown in the figure, the MRT-LBM simulation results agree well with the analytical solution for the whole range of $\varepsilon$. However, the KC equation presents upward discrepancy and downward discrepancy for high and low porosity, respectively.

Further, our SRT-LBM diffusion model is validated by predicting the effective diffusivity in a cubic domain containing a sphere whose diameter is the same as the side length of the cube [19]. For such configuration, the porosity is always 1-π/6. 100×100×100 lattice is employed and the predicted effective diffusivity is 0.3317 in lattice units, in good agreement with the results in [19].

**Acknowledgement**

The authors acknowledge the support of LANL's LDRD Program, Institutional Computing Program, National Nature Science Foundation of China (No. 51406145 and 51136004) and NNSFC international-joint key project (No. 51320105004). Li Chen appreciates the helpful discussions with Doctor Hai Sun from China University of Petroleum, Qingdao.

## Author contributions

L. C. wrote the main manuscript and prepared figures 1-8. J.Y. and L.Z. helped to reconstruct the structures of shales. Q.J.K. and W.Q.T. supervised the theoretical analysis and writing. All authors reviewed the manuscript.

## Additional information

Competing financial interests: The authors declare no competing financial interests.

# Figure captions

Figure 1 Correction factor between apparent permeability and intrinsic permeability predicted by different equations under different Knudsen number

Figure 2 Reconstruction of the shale matrix. (a) A SEM image of a shale rock; (b)3D reconstructed porous structures of shales; (c) pore size distribution of the reconstructed shale; (d) the "transport" (blue) and "dead" (read) portions of void space in the reconstructed shale

Figure 3 Methane concentration distributions in the reconstructed shale

Figure 4 Effects of temperature on effective Knudsen diffusivity

Figure 5 Distribution of pressure and streamline in the reconstructed shale

Figure 6 Correction factor predicted by numerical simulations and corrections in the literature under different pressure

Figure 7 Correction factor predicted by numerical simulations and corrections in the literature under different Kn

Figure 8 Fluid flow and permeability for periodic BBC arrays

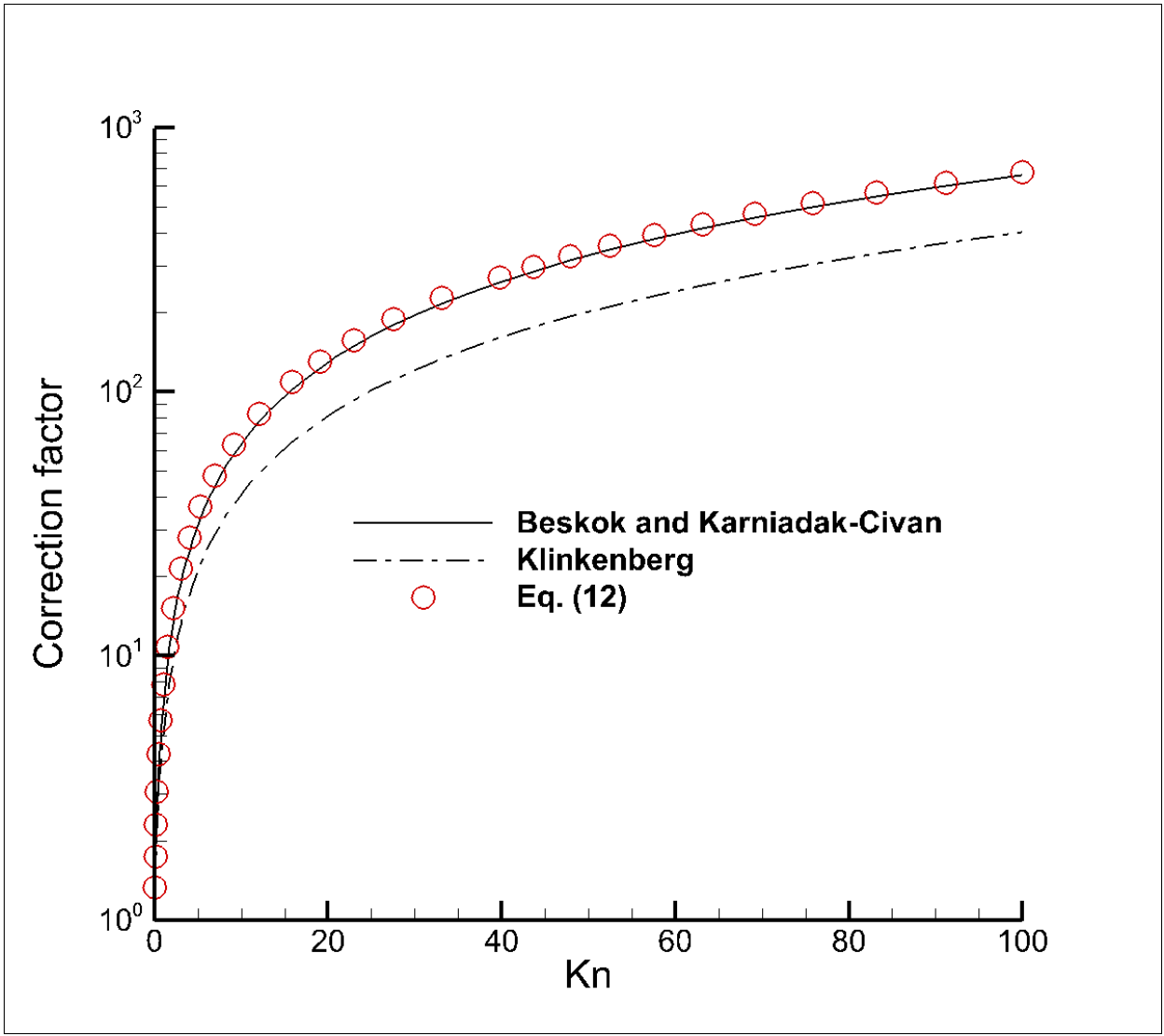

Fig.1

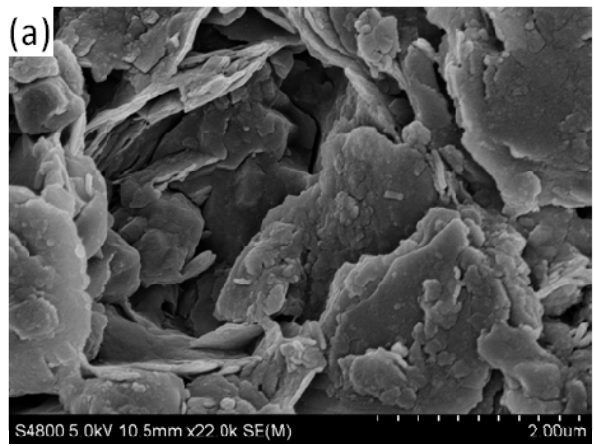
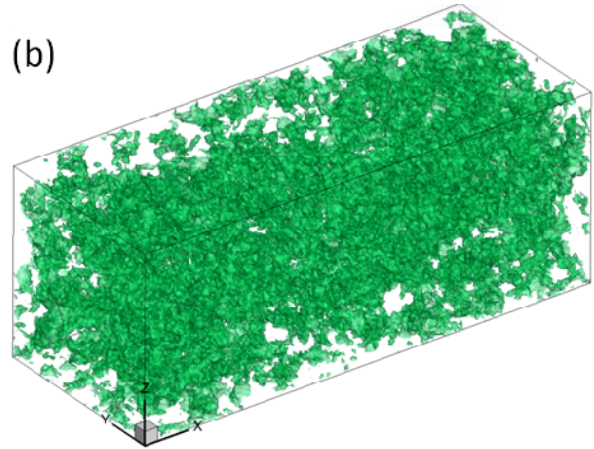
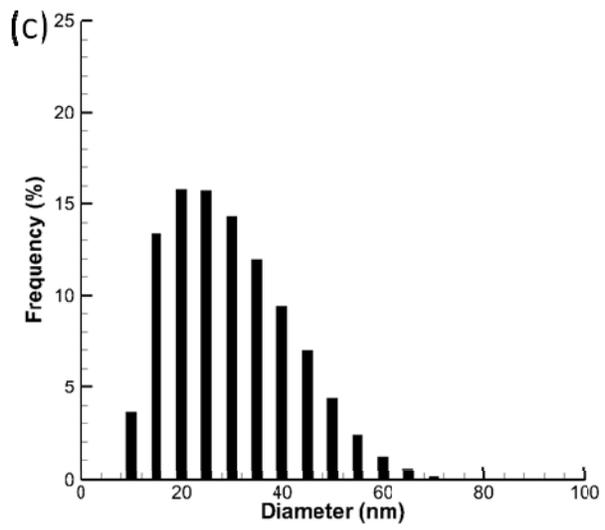
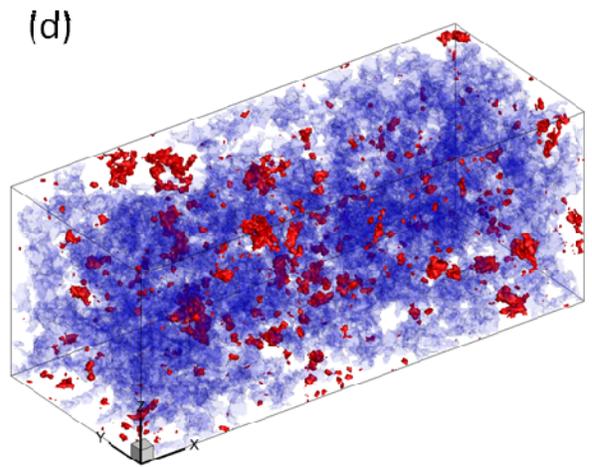

Fig. 2

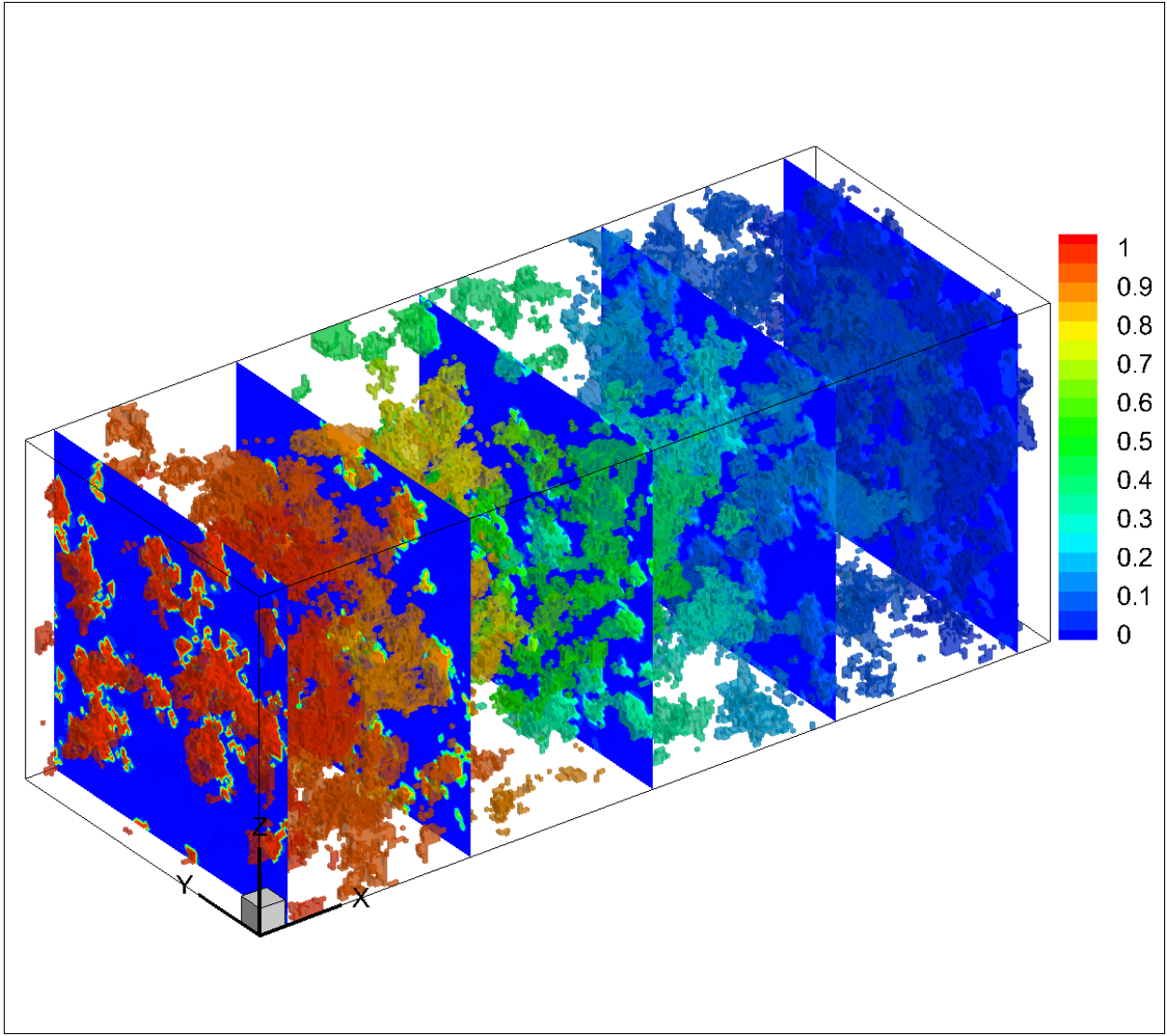

Fig. 3

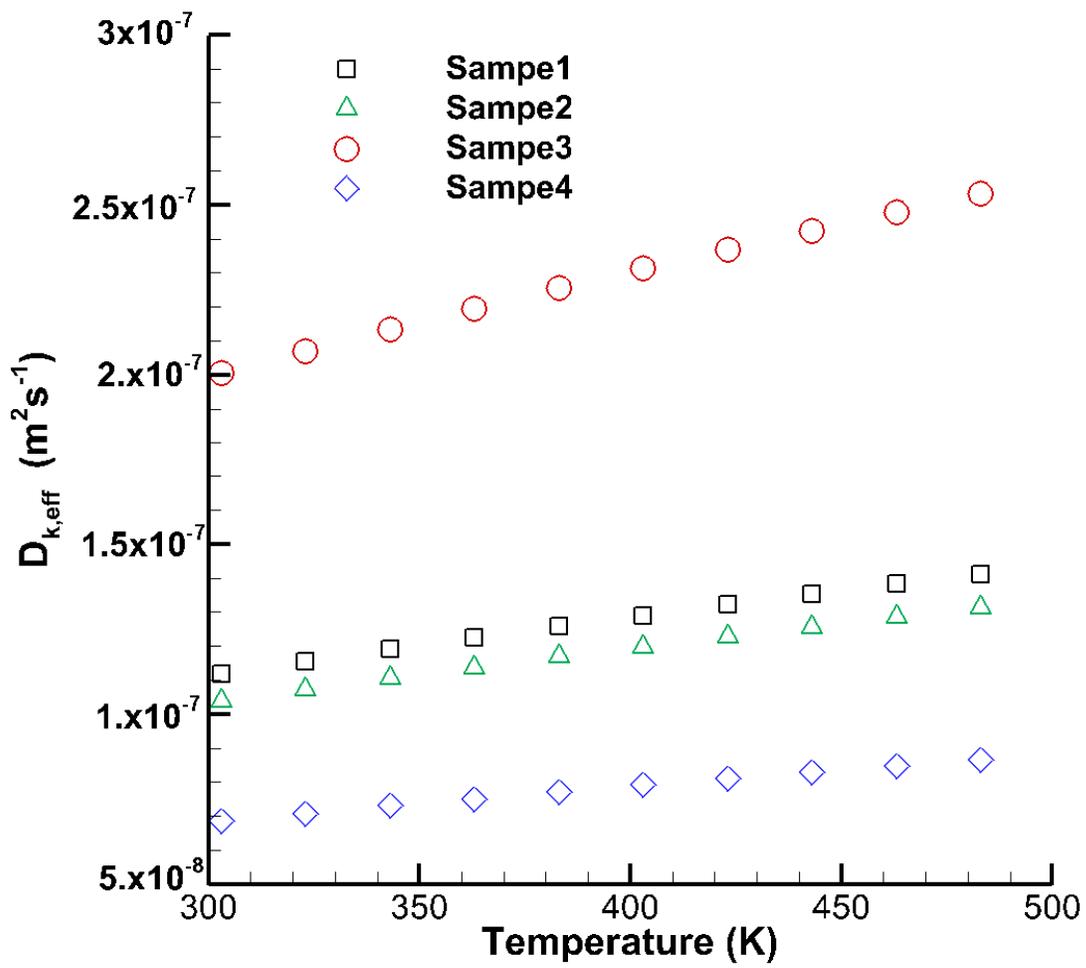

Fig. 4

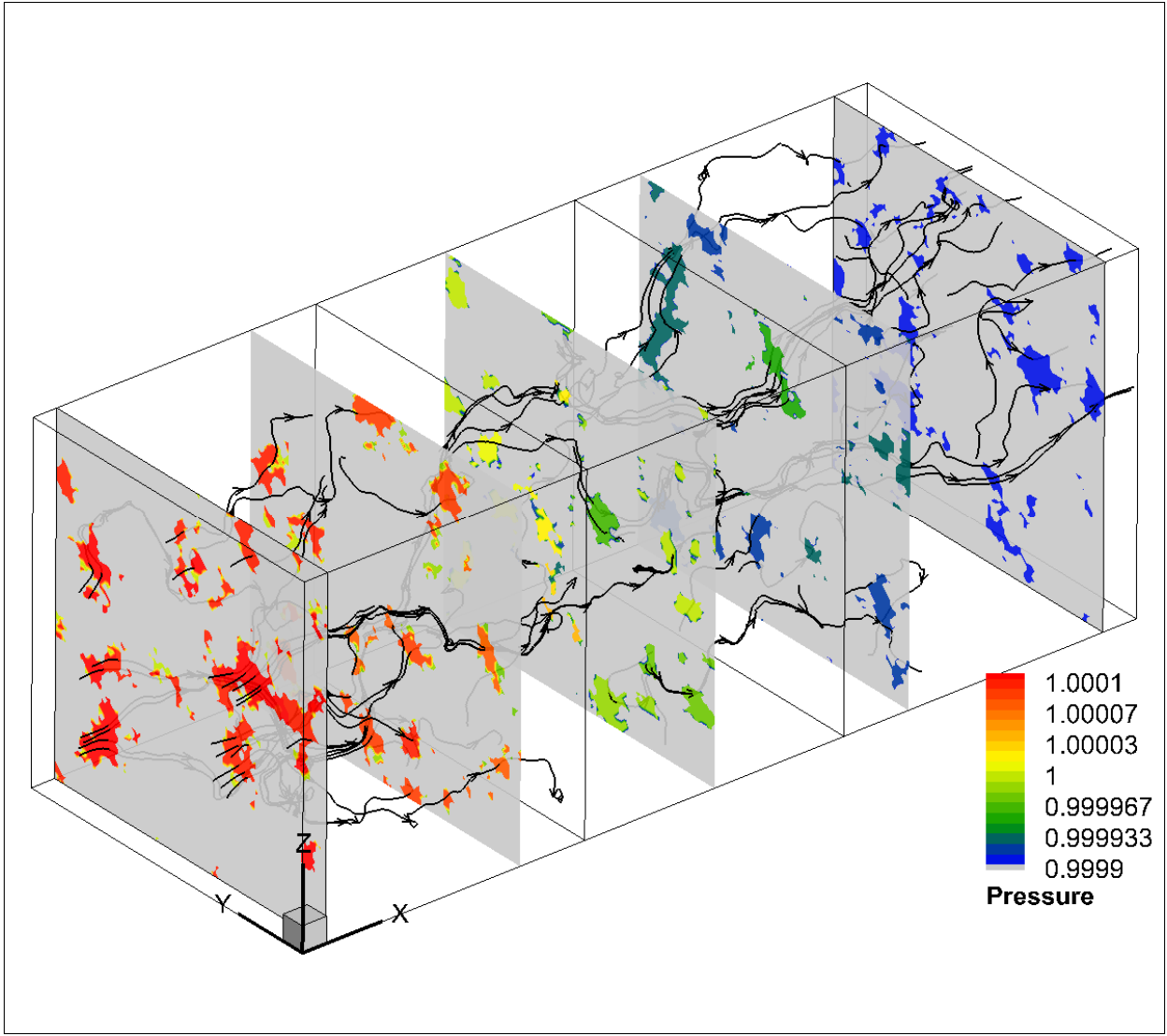

Fig. 5

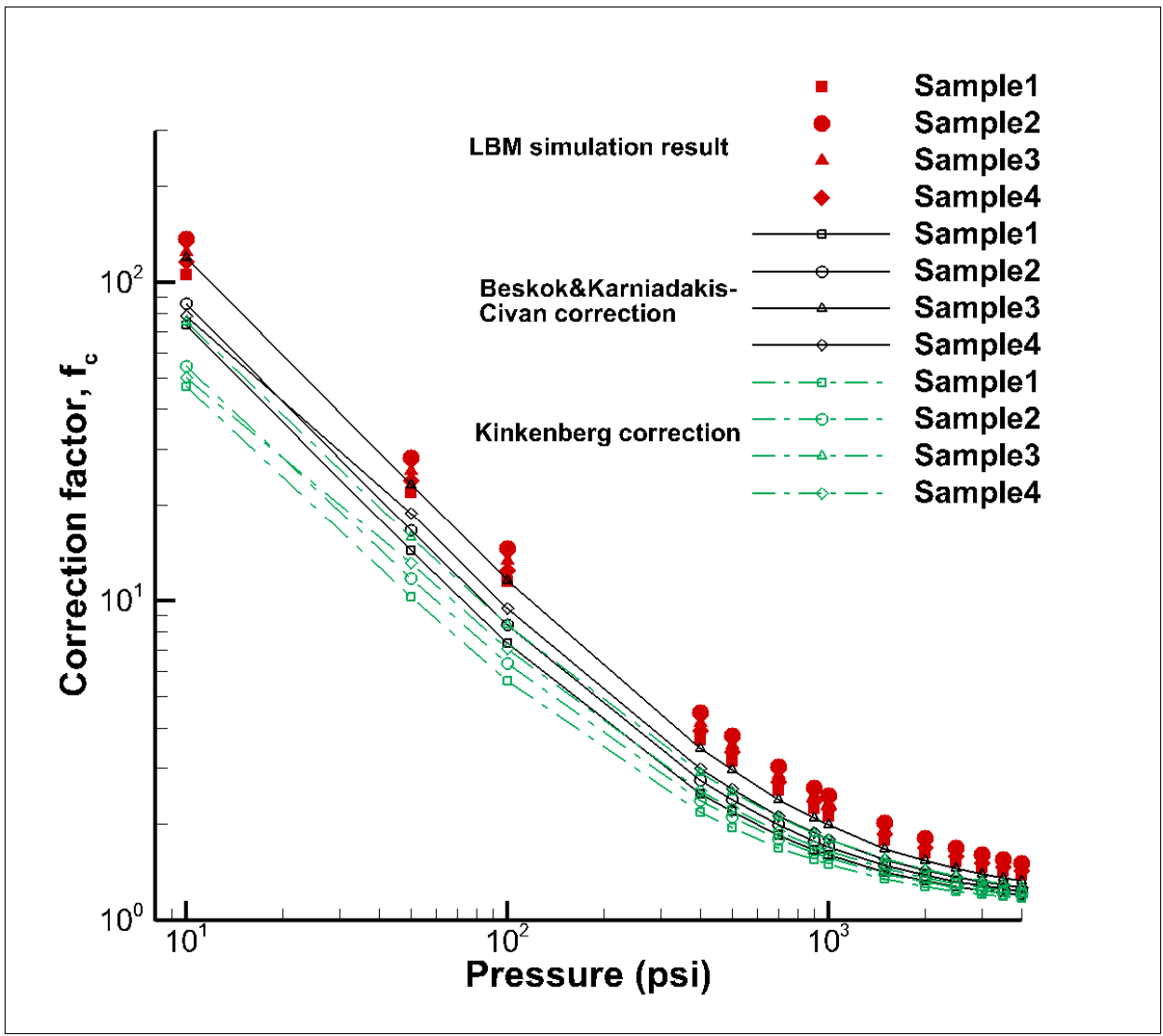

Fig. 6

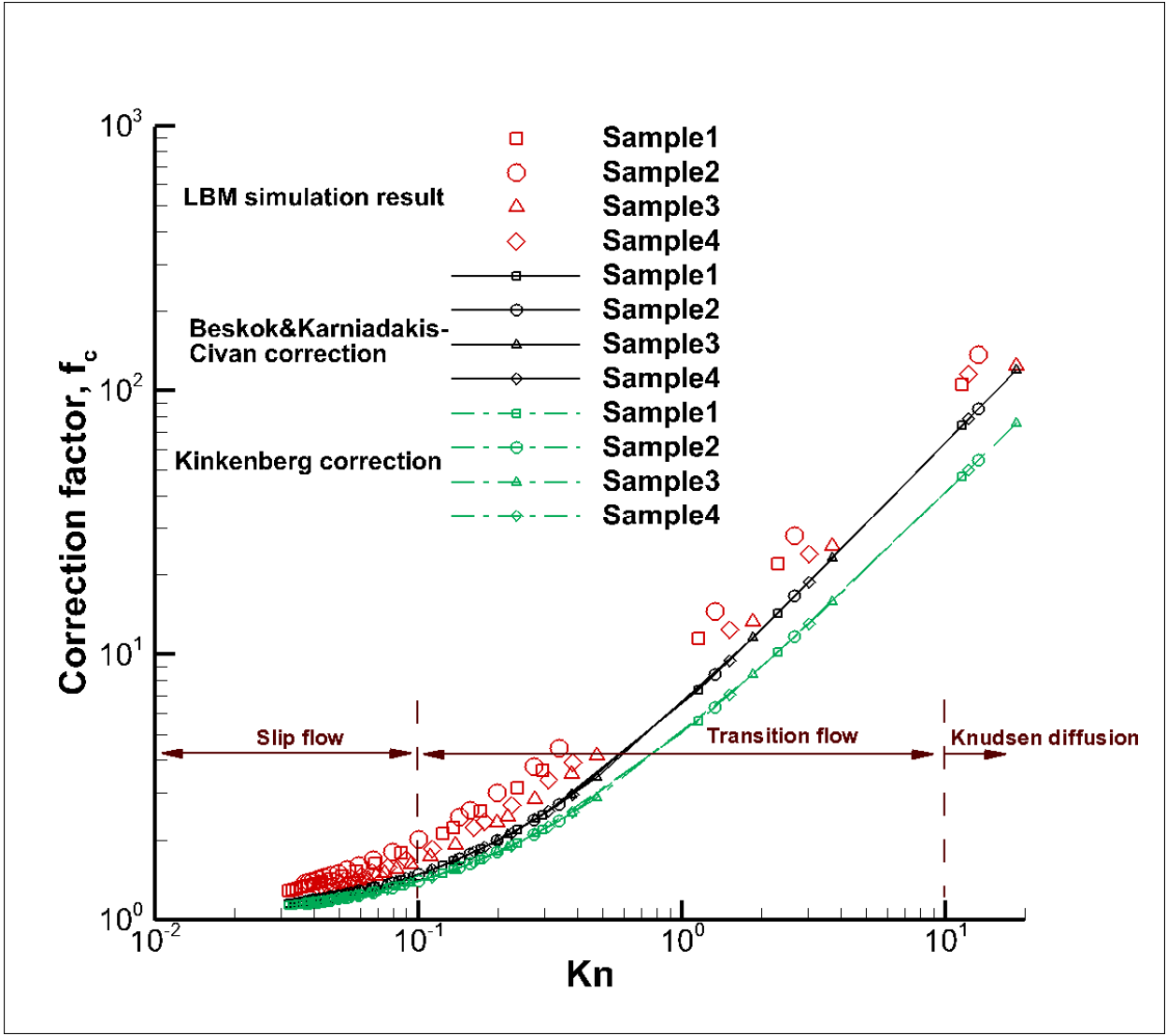

Fig. 7

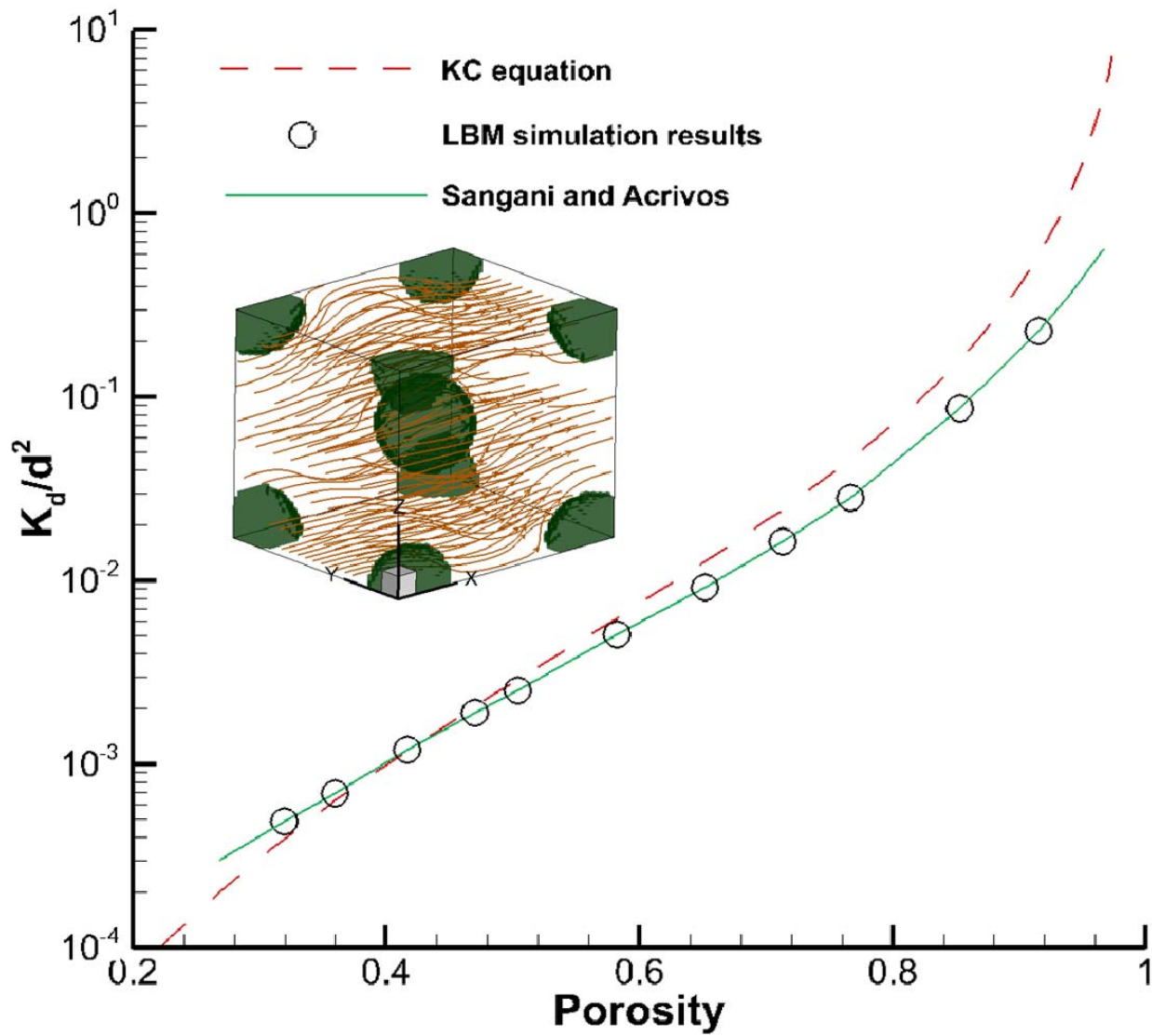

Fig. 8